# Intrinsic doping limitations in inorganic lead halide perovskites

Fernando P. Sabino [1], Alex Zunger [2] and Gustavo M. Dalpian [1]

[1]*Centro de Ciências Naturais e Humanas,*

*Universidade Federal do ABC, 09210-580 Santo André, SP, Brazil.*

[2] *Renewable and Sustainable Energy Institute, University of Colorado, Boulder, Colorado 80309, USA.*

Inorganic Halide perovskites (HP's) of the CsPb$X_3$ ($X$ = I, Br, Cl) type have reached prominence in photovoltaic solar cell efficiencies, leading to the expectation that they are essentially a new class of traditional semiconductors. Peculiarly, they have shown, however, an asymmetry in their ability to be doped by holes rather than by electrons. Indeed, both structural defect-induced doping as well as extrinsic impurity-induced doping strangely result in a unipolar doping (dominantly *p*-type) with low free carriers concentration. This raises the question whether such doping limitations presents just a temporary setback due to insufficient optimization of the doping process, or perhaps this represents an intrinsic, physically-mandated bottleneck. In this paper we study three fundamental Design Principles (DP's) for ideal doping, applying them via density functional doping theory to these HP's, thus identifying the violated DP that explains the doping limitations and asymmetry in these HP's. Here, the target DP are: (i) requires that the thermodynamic transition level induced by the dopants must ideally be energetically shallow both for donors (*n*-type) or acceptors (*p*-type); DP-(ii) requires that the 'Fermi level pinning energies' for electrons $E_{pin}^{(n)}$ and holes $E_{pin}^{(p)}$ (being the limiting values of the Fermi level before a structural defect that compensate the doping forms spontaneously) should ideally be located inside the conduction band for *n*-type doping and inside the valence band for *p*-type doping. DP-(iii) requires that the doping-induced equilibrium Fermi energy $\Delta E_F^{(n)}$ shifts towards the conduction band for *n*-type doping (shift of $\Delta E_F^{(p)}$ towards the valence band, for *p*-type doping) to be sufficiently large. We find that, even though in HP's based on Br and Cl there are numerous shallow level dopants that satisfy DP-(i), in contrast DP-(ii) is satisfied only for holes and DP-(iii) fail for both holes and electrons, being the ultimate bottleneck for the *n*-type doping in Iodine HP's. This study uncovers the hitherto peculiar doping limitations in this class of materials in terms of recognized physical factors.

# I. INTRODUCTION

Doping - the creation of free carriers in a solid by structural and/or chemical substitution - has been a necessary condition for success in many semiconductor technologies based on carrier transport. Yet, the recent advance of Halide Perovskites (HP's) in achieving ~25 % efficient photovoltaic solar cells has occurred despite the rather limited success of doping them.[1–7] For example, in Iodine perovskites, ambipolar doping (both by electrons and by holes, *n*- and *p*-type, respectively) has been reported for $MAPbI_3$, $FAPbI_3$, $CsPbI_3$ just by controlling the ratio of precursors, i.e., the degree of off-stoichiometry.[1–3] However, the achieved carrier density was limited and most of the results show that it does exceed the concentration of $10^{14}$ $cm^{-3}$, which is a very low doping concentration when compared to other traditional semiconductors such as silicon.[4] Bromide and Chloride perovskites such as $MAPbBr_3$, $CsPbBr_3$, and $CsPbCl_3$ show unipolar *p*-type doping with rather low $10^7$ $cm^{-3}$ hole density.[4–7]

Using deliberate *impurity* doping (as opposed to the above intrinsic doping via control of stoichiometry), did not always improve the *n*-type conductivity beyond the intrinsic, stoichiometry control values.[8–11] The substitution of Pb for Bi in $MAPbCl_3$ improved the conductivity by two orders of magnitude (from $2.8 \times 10^{-8}$ to $2.19 \times 10^{-6}$ $\Omega^{-1}.cm^{-1}$),[9] while for the inorganic $CsPbBr_3$ by almost one order of magnitude (from $2.47 \times 10^{-8}$ to $6.75 \times 10^{-7}$ $\Omega^{-1}.cm^{-1}$). After doping, the conductivity of the HP's still lies in the edge of insulator-semiconductor transition.[10] In addition, Bi leads to an increase in the non-radiative process that restricts the possible applications for solar cell devices.[11–14] Transition metals impurities, such as Zr, Hf and Sc were also studied for *n*-type doping in the hybrid $MAPbCl_{3-x}Br_x$ via density functional theory calculations.[15] However, the high formation energy of these defects induce an small shift in the equilibrium Fermi level towards the conduction band, resulting in an equilibrium Fermi energy that lies deep inside the band gap. By consequence the population of electrons in the conduction band will be very small, according to the Fermi-Dirac distribution, and high levels of *n*-type conductivity is unlikely.

The results for *p*-type impurity doping are not different: recent theoretical results showed that Pb substitution by Na, K, Rb, Cu or Ag leads to shallow impurity levels, i.e., part of these impurities will be ionized with thermal energy. Despite that, in HP's, part of the holes created in the valence band will be compensated by structural defects, such as the halogen vacancies or Pb interstitial, and the equilibrium Fermi energy after doping will lie deep inside the band gap, indicating a small doping-induced equilibrium Fermi energy towards the valence band. Because the deep position of the equilibrium Fermi level in the band gap, the population of holes at the valence band will be small according to the Fermi-Dirac distribution and the *p*-type doping in $CsPbCl_3$ is also unlike.[5,16]

In the present paper we inquire whether such doping limitations could present a temporary setback due to insufficient optimization of the process, or perhaps this represents some intrinsic, physically-mandated bottleneck, unlikely to be overcome in these materials at least by conventional near-equilibrium doping methodologies.

The main conclusions of this search can be articulated by posing the three notable theoretical doping bottlenecks and then placing HP's in the context of these "design principles" (DP). To do so, we recall that the doping bottlenecks can be defined by the basic constructs of doping formation energy and transition energies. The intrinsic defects or dopants (*D*) formation energy, $\Delta H_f(D, q)$, in the charge state *q* is given by:

$$\Delta H_f(D,q) = E_{tot}(D,q) - E_{tot}^{bulk} + \sum_i n_i \mu_i + qE_F + \Delta^{(q)}, \qquad (1)$$

where $E_{tot}(D,q)$ is the total energy of the system with defect or dopant *D* in the charge state *q* and $E_{tot}^{bulk}$ is the total energy of the perfect crystal. The defect/dopant is generated by removing (or adding) *n* atoms of chemical specie *i* from (to) the bulk system and adding to (removing from) a reservoir with chemical potential $\mu_i$. Here, $E_F$ is the parametric Fermi energy, i.e., a variable used to compute the formation energy of any defect, and should not be confused with the equilibrium Fermi energy $E_f(T, \{\mu_i\})$, that is calculated according to the neutral charge condition considering all the defects. The former quantity ranges from the valence band maximum with energy $E_v$ to the conduction band minimum with energy $E_c$. Here, $\Delta^{(q)}$ is the finite size correction to Eq. 1. [17,18] The chemical potential $\mu_i$ in Eq. 1 is treated as a variable and can change according to the structural defect or doping and will be discussed in detail in the next section. The thermodynamic transition levels (TTL) $\epsilon(q, q')$ is then defined by the difference of the formation energy of dopant $\Delta H_f(D, q)$ in charge state $q$ and the formation energy of the same defect $\Delta H_f(D, q')$ in charge state $q'$, both calculated in the parametric Fermi level equal to zero ($E_F = 0$), i.e.:

$$\epsilon(q, q') = \frac{[\Delta H_f(D, q) - \Delta H_f(D, q')]}{(q' - q)} . \tag{2}$$

The quantities defined in Eq. 1-2, along with the understanding established earlier on the feedback effect associated with doping,[19–21] can be used to describe the three design principles (DP) that will be required for successful doping, and that are summarized in Fig. 1:

(i) The thermodynamic transition levels $\epsilon(q, q')$ of a dopant must be energetically close to the respective band edges to assure, at least, a small percentage of dopant or defect to be ionized under operating temperatures; this means that the system have defects with "shallow levels" characteristics. It is difficult to define how close the TTL should be from the valence or conduction band, however, here we are going to use as the limit the case of GaN doped with Mg. The calculated ionization energy for this system is around 260 meV, which leads to only 0.1% of Mg ionized at room temperature, despite that, the blue LED based on GaN-Mg works.[22] For energies larger than this, the concentration of ionized impurities decay drastically due to the exponential dependence. Thus, for electron doping, a donor (D) level such as $V_D = \epsilon(+/-)$ in Fig. 1a needs to be close (below in the orange area, or above) the conduction band edge $E_c$, whereas for hole doping the acceptor level $V_A = \epsilon(0/-)$ in Fig. 1b needs to be close (below in the orange area, or above) to the valence band edge $E_v$.

(ii) The n-type (p-type) "Fermi pinning level" $E_{pin}^{(n)}$ ($E_{pin}^{(p)}$) - being the limiting values of the Fermi level reached before a structural "electron killer defect", or "hole killer defect", forms spontaneously - should be located inside the conduction band for electrons (inside the valence band for holes Fig. 1b). Else, doping will be halted by compensation before sufficient carriers are produced. *This pinning is a result of the self-regulating response[19–21] whereby n-type (p-type) doping initially shifts the equilibrium Fermi energy towards the $E_c$ ($E_v$), thereby increasing the likelihood of forming intrinsic structural defects such as acceptors (donors) that compensate the intended n-type (p-type) doping.* In another words, this parameter determine the maximum and the minimum position of the equilibrium Fermi level in the system. And again, if $E_{pin}^{(n)}$ ($E_{pin}^{(p)}$) does not lies in the orange area of Fig. 1 or inside the conduction (valence) band, the maximum possible concentration of electrons (holes) will be very small and the effectively n-type (p-type) doping will be unlikely.

(iii) The shift in the doping-induced equilibrium Fermi energy $\Delta E_F^{(n)}$ towards the conduction band for n-type doping in Fig.1a (shift of $\Delta E_F^{(p)}$ towards the valence band, for p-type doping, Fig.1b) should be sufficiently large for the equilibrium Fermi level to lie close to $E_c$ ($E_v$), i.e., in the orange area. *The magnitude of this doping-induced Fermi level shift can be determined by the energy difference of the equilibrium Fermi level before and after the doping. According to the charge neutrality condition, the equilibrium Fermi level lies close to the region where the lowest formation energy of donor and acceptor are equal, i.e, close to the crossing of $V_X$ and $V_B$ for non-doped system, and close to $V_D$ and $V_B$ ($V_A$ and $V_X$) for n-type (p-type) doping as indicated in Fig. 1. To lead to a relatively large shift in the Fermi level, the donor dopant should have formation energies smaller than the most stable donor structural defect (such as halogen vacancy in HP's). For the acceptor dopant, its formation energies need to be smaller than the most stable structural acceptor defect (such as the Pb and Cs vacancies in HP's).*

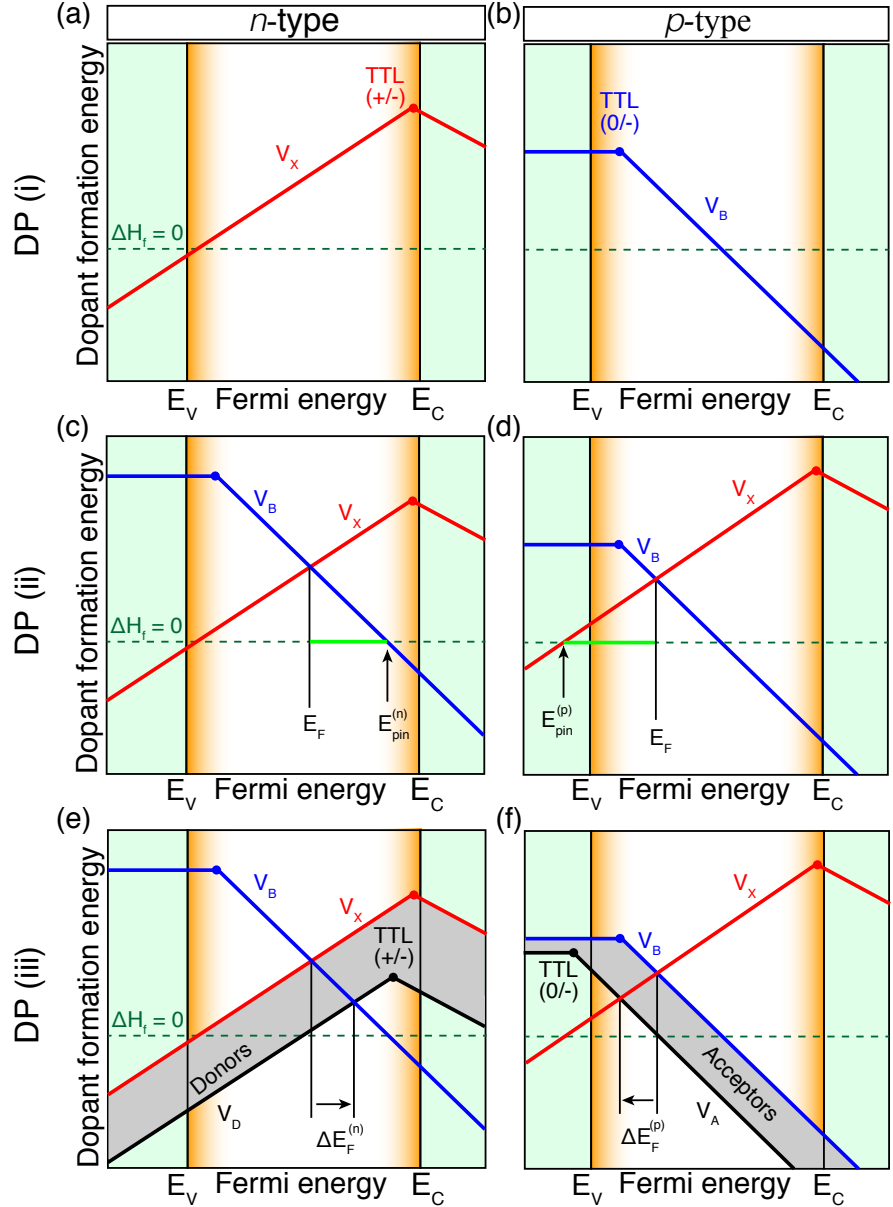

Fig. 1: Schematic representation for the formation energies of structural defects ($V_B$ acceptor and $V_X$ donor) and dopant defect ($V_D$ refers to donors and $V_A$ to acceptors) as function of the parametric Fermi level (ranging from the valence band maximum $E_V$ to the conduction band minimum $E_C$). The three Design Principles that determine successful doping are described in this figure. **DP (i)** – (a) and (b) –The thermodynamic transition levels (represented by a break--solid circle--in the lines) must be shallow, i.e., be located in the orange area. Specifically, the electron producing donor transition (+/-) in n-type doping shown in (a) must be close to the $E_C$, whereas the hole producing acceptor transition (0/-) in p-type doping shown in (b) must be close to $E_V$. **DP (ii)** – (c) and (d) – The Fermi level Pinning $E_{pin}^{(n)}$ ($E_{pin}^{(p)}$) indicated by the vertical arrows must be ideally above $E_C$ (below $E_V$) for n-type (c) (p-type in (d)) doping. This pinning energy is the Fermi energy where intrinsic defects have zero formation energy. The light green horizontal line shows the maximum possible variation of the Fermi level. $E_F$ here indicates the position of the equilibrium Fermi energy according to the charge neutrality condition. **DP (iii)** – (e) and (f) – The doping-induced Fermi energy shift $\Delta E_F^{(n)}$ towards the conduction band for n-type doping (e) (shift of $\Delta E_F^{(p)}$ towards the valence band, for p-type doping, (f)) must be large enough to shift the dopant-induced equilibrium Fermi level close to $E_C$ ($E_V$), i.e., toward the orange area. For the n-type case without the inclusion of donors, the equilibrium Fermi energy can be approximated by the crossing point among the blue and red curves. When donors are inserted, the new equilibrium Fermi energy will be the crossing point among the blue and black lines. Similar analysis can be performed for p-type doping.

In this paper we will illustrate these three Design Principles, showing that they can explain, in a very simple way, the difficulty to dope the HP using structural defects or dopant. Based on the results of CsPb$X_3$ ($X$ = I, Br, Cl), both doping with $p$- and $n$-type will be very difficult to achieve. Despite the doping Design Principles are explored for inorganic halide perovskites in this paper, this strategy belongs to a general idea that can be extended to different perovskite classes and different materials. Therefore, the failure in any conditions (i)-(iii) will result in a bottleneck to achieve the successful doping process in semiconductor HP's, oxides, nitride and others.

## II. THEORETICAL APPROACH AND COMPUTATIONAL DETAILS

*Describing the materials*: Depending on the temperature, halide perovskites CsPb$X_3$ ($X$ = Cl, Br, I) can crystallize in a cubic (*Pm-3m*), tetragonal (*P4mm*) or orthorhombic (*Pnma*) configurations. Recent results have shown that it is possible to tune the transition temperature between these phases by entropy control.[23] Considering these conditions, we calculate all defect and dopant properties in the CsPbBr$_3$ in orthorhombic phase and also we check CsPbI$_3$ and CsPbCl$_3$ in the cubic structure. The theoretical and experimental lattice parameters for all systems are summarized in Table I.

Table I: Theoretical and experimental lattice parameter, band gap and formation enthalpy for CsPb$X_3$ with $X$ = Cl, Br, I.

| System | Lattice parameter (Å) | | | Band gap (eV) | | Formation enthalpy (eV) |
|---|---|---|---|---|---|---|
| CsPb$X_3$ | PBESol | | | PBESol | HSE+SOC | HSE+SOC |
| $X$ = Cl | $a_0$ = 5.62 | | | 2.29 | 2.83 | -8.06 |
| Exp. | $a_0$ = 5.61 [24] | | | | 2.85 [25] | |
| $X$ = Br | $a_0$ = 7.91 | $b_0$ = 8.41 | $c_0$ = 11.64 | 1.82 | 2.27 | -7.46 |
| Exp. | $a_0$ = 8.20 | $b_0$ = 8.25 | $c_0$ = 11.75 [26] | | 2.23 [25] | |
| $X$ = I | $a_0$ = 6.27 | | | 1.64 | 1.72 | -5.86 |
| Exp. | $a_0$ = 6.29 [27] | | | | 1.74 [28] | |

The defects and dopant formation energies in HP's are carried out according to the Eq. 1. Here, we considered only the structural defects with lowest formation energies that include vacancies of Pb, Cs and halogen ($X$ = Cl, Br and I), and the interstitial halogen atoms. All the remaining structural defects are not considered because of the high formation energy, as already demonstrated in previous defect calculations.[4,29] For the dopants, we considered only candidates that can lead to $n$-type doping. The results that support the $p$-type doping are obtained in the literature.[5,16,30,31] Thus, the following dopant elements are calculated only in the Pb site: Al, Ga, In, Sb, Bi, La and Au. The formation energy for dopant on Cs site or interstitial are very high because the large difference in the atomic radius, and therefore were not considered. The thermodynamic transition levels (TTL) of these defects and dopants are calculated with Eq. 2, and these points are mentioned and highlighted in the Results section.

*The DFT used*: To compute the total energy used in the Eq. 1, and all the electronic properties of HP's we used first principles calculations based on density functional theory (DFT) as implemented in Vienna Ab-initio Simulation Package – VASP.[32,33] The projected augmented wave method (PAW)[34,35] was employed to describe the interaction of valence electrons with the ionic cores, considering the following electronic distribution for each chemical specie: Al ($3s^2 3p^1$), Au ($5d^{10} 6s^1$), Br ($4s^2 4p^5$), Bi ($6s^2 6p^4$), Cl ($3s^2 3p^5$), Cs ($5s^2 5p^6 6s^1$), Ga ($4s^2 4p^1$), I ($5s^2 5p^5$), In ($5s^2 5p^1$), La ($6s^2 5d^1$), Pb ($6s^2 6p^2$), Sb ($5s^2 5p^4$). For the exchange and correlation function we employed the Perdew-Burke-Ernzerhof implemented for solid materials (PBESol)[36] and the stress tensor and the atomic forces were minimized with a plane wave energy cutoff of 400 eV. Also we used a **k**-point mesh of 5×5×9 to integrate the Brillouin zone of CsPbBr$_3$ in the primitive orthorhombic cell and the same density for all the remaining systems.

The band gaps of CsPb$X_3$ ($X$ = Cl, Br, I) obtained with PBESol have good agreement to the experimental results as shown in Table 1. It is known, however, that this is an artificial effect. To get an improved description of the electronic properties of halide perovskites it is important to include the spin orbit coupling (SOC) in the calculations.[37] When SOC is added, the band gap is largely underestimated, and can lead to a poor description of defects.[38] To improve the description of the electronic properties, mainly increasing the band gap energy, we employ the hybrid exchange and correlation developed by Heyd-Scuseria-Ernzerhof (HSE)[39,40] in addition to the SOC correction. In the HSE approach, the exchange energy is divided in a short and long range region by a screening parameter $\omega = 0.206 \text{ Å}^{-1}$.[39,40] In the long range, 100% of the exchange energy derives from the PBE functional.[39,40] In the short range, we considered 43% of the exchange energy to come from the Fock operator and 57% from the PBE exchange. These values differ from the default of 25% of exact exchange, however, when combined to the SOC approach, it leads to the correct experimental band gap for all our studied systems (CsPb$X_3$ - $X$ = Cl, Br and I), with a variation smaller than 40 meV as indicated in Table 1.

*Supercell used*: Defects and dopants were computed in a supercell containing 80 atoms and corresponding to a $2\sqrt{2}$x$2\sqrt{2}$x2 repetition of a conventional cubic perovskite cell or a 2x2x1 repetition of the orthorhombic cell. We also performed the calculation for a supercell of $2\sqrt{2}$x$2\sqrt{2}$x4 (2x2x2) of the conventional cubic (orthorhombic), which contains 160 atoms and the results for defect formation energies differ by less than 0.15 eV when compared to the smaller cell. In order to precisely simulate the cubic structure, it is important to take into account its polymorphous configuration.[41] Under this consideration, the cubic cell shows a cubic Bravais lattice by the average of the atomic positions. In this configuration, the local chemical environment for each octahedra (formed by Pb bonded with six halogen atoms) is allowed to distort, becoming locally similar to the orthorhombic configuration. Because of this characteristic, we do not expect a large difference in the formation energy of defects when comparing orthorhombic and cubic polymorphous cells even with the difference in the lattice parameters.

*Chemical potential and phase diagram*: As mentioned before, the chemical potentials in Eq. 1 are treated as a variable and they must satisfy the stability condition for the formation enthalpy of each perovskite system, which is given by:

$$\mu_{\text{Cs}} + \mu_{\text{Pb}} + 3\mu_X = \Delta H_f(\text{CsPb}X_3), (X = \text{Cl, Br, I}), \quad (3)$$

where $\Delta H_f(\text{CsPb}X_3)$ is the formation enthalpy for each system CsPb$X_3$ that is shown in Table 1. To avoid the precipitation of competing phases, the sum of the chemical potential of each element should not overcome the formation enthalpy of these competing phases. Therefore, they must follow the inequations of instability. Using CsPbI$_3$ as an example we have:

$$\mu_{\text{Cs}} + \mu_{\text{I}} < \Delta H_f(\text{CsI}),$$
$$\mu_{\text{Cs}} + 3\mu_{\text{I}} < \Delta H_f(\text{CsI}_3),$$
$$\mu_{\text{Pb}} + 2\mu_{\text{I}} < \Delta H_f(\text{PbI}_2), \quad (4)$$

where $\Delta H_f(\text{CsI})$, $\Delta H_f(\text{CsI}_3)$, $\Delta H_f(\text{PbI}_2)$ are the formation enthalpies for the most stable competing phases for CsPbI$_3$. If we consider CsPbBr$_3$, the same procedure is applied, with the competing phases being PbBr$_2$, CsBr, CsBr$_3$, Cs$_4$PbBr$_6$ and CsPb$_2$Br$_5$, while for CsPbCl$_3$ we considered the following competing phases: PbCl$_2$, CsCl, Cs$_2$PbCl$_6$, CsPb$_2$Cl$_5$. With this condition we can determine the stability region (stability triangle) with respect to the chemical potentials for each perovskite as shown in Fig. 2. The full stability triangles are shown in the supplementary material Fig. S4.

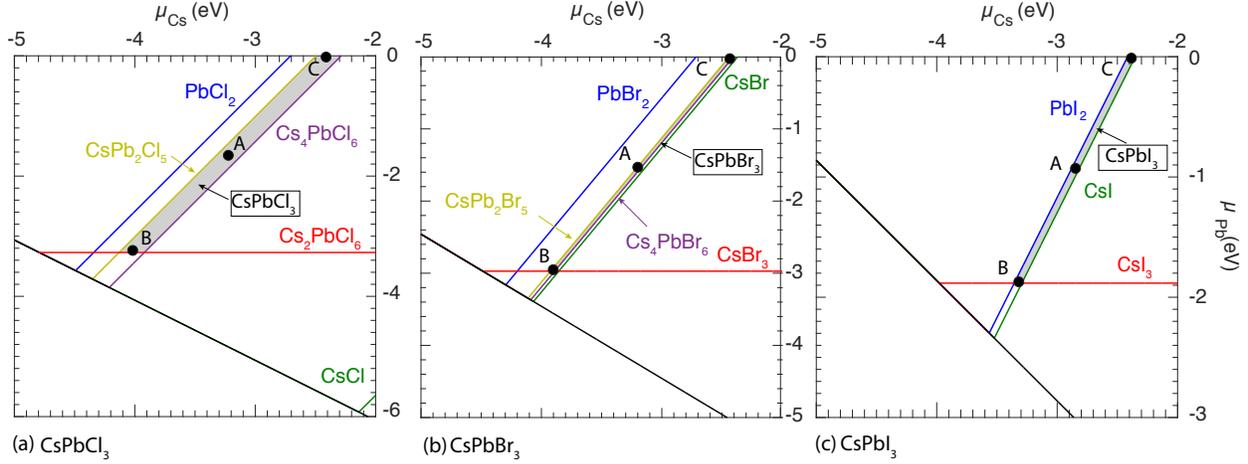

Fig. 2: Stability triangle for CsPb$X_3$ (a) $X$ = Cl, b) $X$ = Br and c) $X$ = I). The perovskite phase will form for chemical potentials of Pb and Cs within the B-C line (shaded region), which represent the closest condition for halogen rich and halogen poor, respectively. Point A represents the intermediate condition, and it is determined as a condition that leads to stability of all extrinsic dopants in the perovskites analyzed here.

Each chemical potential can vary between the B-C line in Fig. 2, and these points represent the halogen rich and poor conditions respectively. However, when the dopant is incorporated in the system, the thermodynamic stability condition should avoid the precipitation of the doping precursor or the formation of other compounds. If we consider the precursors of the dopants in the halogen base (for example to dope CsPbI$_3$ with Bi, the precursor used is BiI$_3$) the chemical potential of halogen is limited by the inequation:

$$\mu_{\text{Bi}} + 3\mu_{\text{I}} < \Delta H_f(\text{BiI}_3), \quad (5)$$

where $\Delta H_f(\text{BiI}_3)$ is the formation enthalpy of the precursor used to dope the halide perovskite with Bi. This equation demonstrates that we have low incorporation of the dopant close to the halogen poor condition (point C). Similar limitation is also observed for all the other elements used to dope the HP's in this study. In the supplementary material Figure S1, we shown the possible chemical potentials to avoid the precipitation of the doping precursor for each element. Therefore, considering all the possible dopant elements, we choose an intermediate point, named as point A in the phase diagram of Fig. 2, that leads to a stability to incorporate the dopant element in all CsPb$X_3$ ($X$ = Cl, Br, I) systems. This intermediate condition also represents the stochiometric precursors ratio in the system.

*DP properties calculations*: Considering the formation energy determined by Eq. 1 for the structural defects and dopants, and also the chemical potential in the stability triangle, we can compute the Fermi pinning levels for *p*- and *n*-type doping, i.e., $E_{pin}^{(p)}$ and $E_{pin}^{(n)}$ in the HP's. These parameters indicate the asymptotic values of the Fermi level reached before a structural "electron killer defect" such as acceptor ("hole killer defect", such as a donor) form spontaneously ($\Delta H_f(D, q) < 0$). In the HP's studied here, the Fermi pinning levels for *p*-type ($E_{pin}^{(p)}$) is determined when the formation energy of halogen vacancy is equal to zero ($\Delta H_f(V_X, +) = 0$) at the halogen richest condition, close to the B point of Fig. 2. For *n*-type Fermi pinning levels, $E_{pin}^{(n)}$ is determined when the acceptor Pb vacancy has formation energy equal to zero ($\Delta H_f(V_{\text{Pb}}, -2) = 0$) at the Pb richest condition (halogen poor), close to the C point of Fig. 2. For an effective doping in a material, it is desirable that $E_{pin}^{(n)}$ ($E_{pin}^{(p)}$) lies in the conduction (valence) band, i.e., it is possible to populate the conduction (valence) band with a certain amount of electrons (holes) before the intrinsic defect start to compensate it. If $E_{pin}^{(n)}$ ($E_{pin}^{(p)}$) lies deep inside the band gap, the *n*-type (*p*-type) of doping will be unlikely in system.

The doping-induced equilibrium Fermi energy $\Delta E_F^{(n)}$ and $\Delta E_F^{(p)}$ are calculated based on the difference of equilibrium Fermi level before and after doping. The equilibrium Fermi level can be determined by the charge neutrality condition, which depends on the structural defects and dopant formation energies, their density in the cell and the density of states of valence and conduction bands. In a first approximation, this calculation leads to an $E_F$ close to the condition where the formation energy of the most stable donor and acceptor, among all the structural defects and dopants, are equal. In the plot of Formation energy as function of the parametric Fermi level, the equilibrium Fermi level is located close to the crossing of the donor and acceptor with lowest formation energies, as indicated in Fig. 1. For the HP's studied here, $\Delta E_F^{(n)}$ is determined by the difference of the parametric Fermi level at the crossing of halogen vacancy and Pb or Cs vacancy (non-doped system) and the parametric Fermi level at the crossing of donor dopant with Pb or Cs vacancy (doped system). In this work we do not calculate $\Delta E_F^{(p)}$ because we only considered dopants that are candidates for *n*-type doping. Values for *p*-type doping can be extracted from previous theoretical results in the literature.[5,16,30,31]

## III. Applying the 3 Design principles (DP) for successful doping to Halide Perovskites: General observations

Here we use the 3 design principles (i)-(iii) explained in the Introduction and Fig. 1 to analyze our DFT results **why** is it impossible to effectively dope halide perovskites of CsPb$X_3$ (*X* = Cl, Br, I) based on physical arguments and analysis of the formation energy of structural defects and dopants. The results that support this discussion are shown in the next section for the *n*-type doping, while recent results from the literature support the *p*-type doping analysis.[5,16,30,31] It is important to note that our argument is valid for electronic doping, and therefore we are not considering ionic conductivity.

**Regarding the DP (i):** *The thermodynamic transition levels $\epsilon(q,q')$ of a dopant must be energetically close to the respective band edges ("shallow level") to assure its ionization under operation conditions.* For such shallow TTL, the thermal energy is sufficient to promote electrons from the dopant level to become a free electron in the CB, and this impurity is a good candidate for *n*-type doping. It is difficult to define what is the energy range of a "shallow" defect when the TTL lies in the band gap. However, here we consider the maximum energy of 260 meV that leads to 0.1% of ionized structural defects or impurities at room temperature, as calculated for Mg doped GaN and explained before.[22] The concentration of ionized defects or impurity has an exponential dependency with the energy, and therefore values larger than 260 meV will lead to a negligible charge concentration in valence or conduction band. Most of the simple structural defects in halide perovskites are shallow, include vacancies of halogen ions (donors), as well as vacancies of the cations (acceptors) Pb, Sn, Cs and MA, as demonstrated by DFT calculations in the literature and in the second part of this paper.[1–3,5,15] Because of these characteristics, several experiments have been trying, without success, to dope HP only with the structural defects by changing the chemical potential (or the precursors ratio in the synthesis).[4] In addition, substitution of a Pb atom by a donor dopant, for example Bi or Sb, also leads to shallow TTL lying in the band gap, closer to the CBM, and therefore should be considered good candidates for *n*-type doping. However, the HP conductivity increases (when it increases) by only one or two orders of magnitude when compared to the intrinsic material, indicating low levels of *n*-type doping. Similarly, the substitution of Pb by atoms from family IA, such as Na, K, Rb or transition metals like Cu or Ag leads to shallow TTL for *p*-type doping in most of HP's.[5,16] In this condition the improvement of the carrier concentration is still far from satisfactory for a *p*-type doping. Fig. 1 depicts a schematic formation energy diagram for a condition of shallow intrinsic and extrinsic defects in both kinds of doping. In this figure, V$_X$ represents a shallow structural donor (transition level close to the CBM) and V$_B$ represents a shallow structural acceptor (transition level close to the valence band minimum). If we consider only the argument of shallow TTL to dope a HP's, all these examples should lead to successful doping. Nevertheless, this is not experimentally observed.[4]

***Regarding the DP (ii):*** *The n-type (p-type) Fermi pinning level $E_{pin}^{(n)}$ ($E_{pin}^{(p)}$) should be located inside the conduction band; Fig 1a (inside the valence band; Fig. 1b).* The dilution of an impurity in HP increases the carrier density, and this extra charge moves the Fermi level toward the conduction (donor) band or valence band (acceptor). The different position of the Fermi level leads to a different thermodynamic equilibrium condition, decreasing the formation energy of few intrinsic defects that tends to kill the extra charge introduced by the dopant. For example, when extrinsic donor dopant is introduced in CsPb$X_3$ ($X$ = I, Br and Cl), it will shift the Fermi energy upwards and, consequently, the formation energy of an intrinsic acceptor, e.g. Pb vacancy, will decrease. This will lead to a larger population of these Pb vacancies, that will capture the extra electrons provided by the donors. Acceptor impurities can show the same behavior, but now the halogen vacancies are the main source of hole killers. However, when the formation energy of a structural defect becomes negative (spontaneously form), the charge introduced by the dopants will be automatically killed by the structural defects. This condition determines the *Fermi pinning level* [19,21,42] of the system and is represented by $E_{pin}^{(n)}$ and $E_{pin}^{(p)}$ for *n*- and *p*-type doping, respectively, as shown schematically in Fig. 1. For an effective doping, the pinning of the Fermi level must lie in the valence band (VB) and conduction band (CB) for *p*-type and *n*-type doping, respectively. If the pinning of the Fermi level lies deep inside the band gap, the intrinsic defects will compensate the charges introduced by the impurities before the Fermi level reach the VB or CB, and therefore the doping will be unsuccessful. In other words, the pinning of the Fermi level is the parameter that determine the possible maximum and minimum position (taking the $E_V$ as reference) of the equilibrium Fermi level in the system. For CsPb$X_3$, we calculate the pinning of the Fermi level (second part of the paper) for *p*- and *n*-type. In all of these materials, the $E_{pin}^{(p)}$ lies in the VB, with energy $E_v$ - 0.11 eV, $E_v$ - 0.56 eV and $E_v$ - 0.47 eV for $X$ = I, Br and Cl, respectively. In contrast, $E_{pin}^{(n)}$ lies in the band gap with energy, $E_c$ - 0.27 eV and $E_c$ - 0.55 eV for CsPbBr$_3$ and CsPbCl$_3$, respectively. This explains why most of perovskites based on Br and Cl have a tendency for unipolar *p*-type doping.[5–7] In CsPbI$_3$, $E_{pin}^{(n)}$ is at the edge of the conduction band minimum, and in principle, *n*-type doping could be achieved with a low magnitude. This behavior is observed in HP's based on I that shown a bipolar type doping with low intensity.[1,2] Nevertheless, even for perovskites with pinning of Fermi level in the VB, the *p*-type doping is never achieved as demonstrated for CsPbCl$_3$.[5] To understand this, we need to check the third *DP* necessary for doping to be successful.

***Regarding the DP (iii):*** *The shift in the doping-induced equilibrium Fermi energy $\Delta E_F^{(n)}$ towards the conduction band for n-type doping in Fig. 1a (shift of $\Delta E_F^{(p)}$ towards the valence band, for p-type doping, Fig. 1b) is sufficiently large.* Since the pinning of the Fermi level for *n*-type doping $E_{pin}^{(n)}$ in systems based on Br and Cl lies in the band gap, the shift promoted by extrinsic dopant such as Bi or Sb will not lead to a considerable improvement in the conductivity. For CsPbI$_3$, the pinning of the *n*-type Fermi level is close to the conduction band, however the *doping-induced Fermi energy shift* $\Delta E_F^{(n)}$ *is* very small, of the order of 0.22 eV, which represent less than 13% of the band gap energy (band gap shown in Table 1). This explain why the *n*-type doping is also very difficult to achieve in HP's based on I. The same argument is used to prove why *p*-type doping in CsPbCl$_3$ is hard to be achieved.[5] The *p*-type pinning energy $E_{pin}^{(p)}$ in CsPbCl$_3$ lies in the valence band, however the *doping-induced Fermi energy shift* $\Delta E_F^{(p)}$ *towards the valence* band after doping with acceptor impurities Na, K, Rb, Cu or Ag is still far away from the valence band, lying deep in the band gap.[5] The $\Delta E_F^{(p)}$ is very small and makes the *p*-type doping unsuccessful.

The DP (ii) and (iii) are a consequence of the valence and conduction band position with respect to the vacuum. When the atomic radius of halogen decreases, i.e., goes from I to Cl, the position of $E_v$ tends to become deeper in energy, increasing the ionization potential for the HP's, as shown in Fig. S3 in the supplementary material.[43] High ionization potential increases the difficulty to find shallow acceptors that leads to a large $\Delta E_F^{(p)}$ and therefore the *p*-type doping becomes unlikely to be achieved.

For the *n*-type, the decrease in the halogen radius also decrease the electronic affinity of the halide perovskite, and the $E_c$ becomes closer to the vacuum level for Cl when compared to I or Br, as shown in Fig. S3 in supplementary material.[43] Thus, the difficulty to find shallow donors that leads to large $\Delta E_F^{(n)}$ increases and the process to achieve the *n*-type doping becomes unlikely.

In the next section, through DFT calculations, we determine the Fermi pinning levels and $\Delta E_F^{(n)}$ and $\Delta E_F^{(p)}$ for CsPb$X_3$ (*X* = I, Br and Cl), that is the base for our DP already demonstrated. In addition, we discuss the details of atomic configuration and electronic properties for each defect and dopant.

## IV. Structural defects and extrinsic impurities as dopants

### A. Structural doping defects

For *n*-type doping, only the structural defects with small formation energies will be considered, i.e., $V_X$, $V_{Cs}$, $V_{Pb}$ and $X_i$. The anti-site defects (Pb$_{Cs}$, Pb$_X$, $X_{Pb}$, $X_{Cs}$, Cs$_{Pb}$, Cs$_X$) or Cs$_i$ and Pb$_i$ have higher formation energies or contribute only for a condition that leads to a possible *p*-type doping, as demonstrated in previous DFT calculations.[5] Therefore, these defects are not considered here. All the formation energy for the most stable intrinsic defects are shown in Fig. 4.

*Halogen vacancy*: For the halogen vacancy in the neutral charge state ($V_X^0$), when one *X* atom is removed from the CsPb$X_3$ perovskite structure, two *X*-Pb bonds are broken, and one unpaired electron is created. This configuration results in an occupied Kohn-Shan (KS) vacancy level that lies inside the band gap and an unoccupied state in resonance in the CB. In addition, there is a tendency for Pb atoms in the vicinity of the vacancy to move inward (outward) the vacancy site, resulting in variation of the Pb-Pb distance of −6.76% and −7.80% (9.11%) for *X* = Cl and Br (I), respectively, when compared to the system without the defect. This effect can be observed by the atomic configuration shown Fig. 3 with its respectively charge density for the defect level in CsPbI$_3$ (the other systems have results similar to this).

When the unpaired electron is removed from the previous condition, resulting in the configuration $V_X^{+1}$, the vacancy level which was occupied in the $V_X^0$ charge state becomes unoccupied and its energy moves to inside the CB. This electronic configuration results in a Pb movement outward the vacancy, increasing the distance Pb-Pb by 4.98%, 5.03% and 9.63% for *X* = Cl, Br and I respectively, and compared to the system without vacancy. This configuration is also shown in Fig. 3 for CsPbI$_3$.

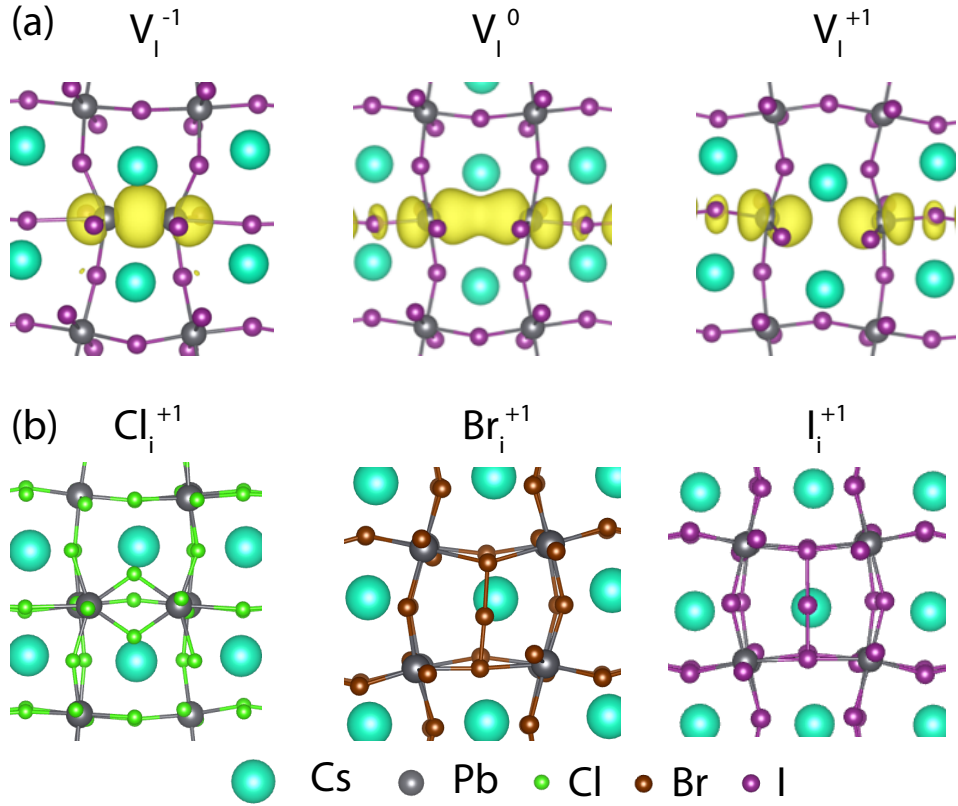

Fig. 3: a) Atomic configuration and charge density for an Iodine vacancy in CsPbI$_3$ in -1, neutral and +1 charge state. The charge density is related to the eigenstates located inside the bandgap. b) Atomic structure for the most stable configuration of the halogen interstitial atom in the +1 charge state for CsPbCl$_3$, CsPbBr$_3$ and CsPbI$_3$. The chemical elements and their respective colors are shown in the bottom of the Figure.

On the other hand, when an extra electron is added to the neutral vacancy forming the $V_X^{-1}$ charge state, the vacancy level of $V_X^0$ that was unoccupied becomes occupied, and results in two levels inside the band gap (not degenerated due to SOC). This electronic configuration results in a movement of Pb inwards towards the vacancy, as shown in Fig. 3, and there is a tendency for Pb-Pb dimer formation with the distance between these atoms decreasing by −33.10%, −35.36% and −42.49% for $X$ = Cl, Br and I, respectively.

Considering all the possible charge states for the halogen vacancy, we find that $V_X^q$ with $q$ = 0 is a high energy defect and is unstable with respect to the -1 and +1 charge state in CsPb$X_3$ ($X$ = Cl, Br, I) according to the formation energy shown in Fig. 4. The calculated transition energy from Eq. 2 indicates that the halogen vacancy is a negative Coulomb interaction (negative U), i.e. there is a tendency for a direct transition from positive +1 to negative -1 charge state in agreement with previous results in the literature.[29] This is because the energy gain owing to structural relaxation of $V_X^{-1}$ suppresses the electron-electron repulsion energy when compared to neutral charge state.

For CsPbI$_3$, the TTL for the $X$ vacancy (+/-) occurs deep inside the conduction band, while for CsPbBr$_3$ it is at the edge of the $E_c$, as shown in Fig. 4. In both situations, the $V_X$ can be considered as a shallow donor. In contrast, for CsPbCl$_3$ the TTL (+/-) for $V_X$ lies at $E_c$ - 0.30 eV, which is nearly a deep donor.

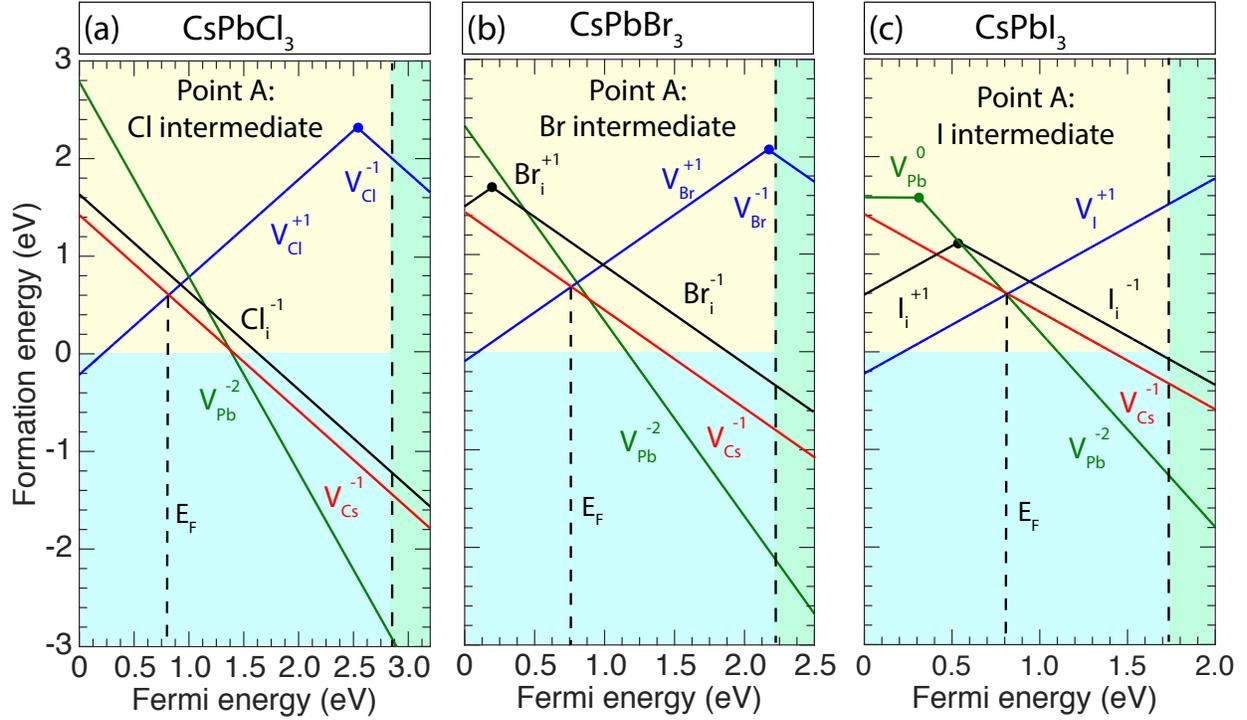

Fig. 4: Formation energy for structural defects as a function of the Fermi level for CsPb$X_3$ with $X$ = (a) Cl, (b) Br and (c) I, respectively. The formation energies were calculated using the chemical potentials related to points A in the phase diagram of Fig. 2. The green area to the right indicates the limit of the conduction band minimum (note different band gap scales for each compound). The blue and yellow regions represent the spontaneous and non-spontaneous formation of defects (negative and positive formation energies). The equilibrium Fermi level was determined by the crossing of the donor and acceptor defects with the lowest formation energy.

*Cesium vacancy*: Being highly ionic, the Cs vacancy does not lead to dangling bonds in the structure. Instead, they leave one hole in the valence band which is derived from halogen $X$ $p$-orbitals.[44] Different from the oxides perovskites, such as SrTiO$_3$, the hole does not tend to form a polaron.[45]

In all HP's studied here, we observed that the Cs vacancy has the -1 ($V_{Cs}^{-1}$) charge state with the lowest formation energy. Other tested charge states (e.g., neutral and +1) have higher formation energies. Because of that, $V_{Cs}$ is considered a shallow acceptor defect and act as a source of electron compensation in CsPb$X_3$ ($X$ = Cl, Br, I). The formation energy of $V_{Cs}^{-1}$ has negative values for a wide range of $E_F$, especially close to the conduction band. This result is independent of the halogen chemical specie despite the intensity is larger for $X$ = Cl. The main reason for that is associated to the larger (smaller) ionic potential (electron affinity) of CsPbCl$_3$ that pushes the $E_v$ ($E_c$) to deeper (shallow) energies when compares to the vacuum level, as shown in Fig. S3 in the supplementary material.

*Lead vacancy*: When the Pb vacancy is introduced in CsPb$X_3$ ($X$ = Cl, Br, I), six Pb-$X$ bonds are broken and two holes are left in the valence band. Similar to $V_{Cs}$, those two holes do not tend to localize, forming small polarons. The distance of two neighboring halogen atoms is so large that the $X$-$X$ bond is avoided for $X$ = Br and Cl. Nevertheless, the largest atomic radius of I (compared to Cl and Br) results in a larger interaction of the halogen atoms leading to a formation of a trimer I-I-I inside the $V_{Pb}$ by a dislocation of one I atom. This atomic configuration is observed only in the neutral charge state, $V_{Pb}^{0}$, and for $X$ = I, as shown in Fig. 4 and also demonstrated previously in the literature.[29] For all the remaining conditions, Pb vacancy is stable only in the -2 charge state, $V_{Pb}^{-2}$, when compared to the other possible charges (e.g., neutral and -1). Because of this characteristic, $V_{Pb}$ is a shallow acceptor and, with $V_{Cs}$, can be considered a source of electron compensation in HP's.

*Interstitial halogen atom*: The other most stable intrinsic defect shown here is the halogen interstitial atom. When this extra halogen atom is added to the system, it can occupy different sites, forming, for example, trimers of halogen atoms (*X-X-X*),[29,46] or modifying two adjacent Pb local chemical environments, increasing the number of bonds with halogen atoms from six to seven as shown in Fig. 3. We tested both configurations for all CsPb$X_3$ (*X* = Cl, Br, I) systems, and observed that the charge state and the atomic radius play an important role for this defect.

The neutral charge state has a high formation energy and is not stable when compared to the +1 or -1 configurations for all halogen chemical species, as shown in Fig. 4. In CsPbI$_3$ and CsPbBr$_3$, both +1 and -1 charge states have the lowest formation energies, with a predominance of +1 close E$_v$, and indicating a negative Coulomb interaction, as also observed for the halogen vacancy. However, the negative charge has the lowest formation energy for CsPbCl$_3$, and therefore is the most stable charge state when compared to the neutral and +1. The transition (+/-) occurs 0.59 and 0.19 eV above the E$_v$ for CsPbI$_3$ and CsPbBr$_3$, respectively. Therefore, this defect can act with different behavior according to the halogen specie. For CsPbCl$_3$ and CsPbBr$_3$, $X_i$ is a shallow acceptor, and a possible source of electron compensation. For CsPbI$_3$, $X_i$ can be considered a deep donor or a deep acceptor, owing to the negative Coulomb interaction, and is the intrinsic defect that has a deep behavior with the smallest formation energy.

The shift of the thermodynamic transition level (+/-) towards the E$_v$ when the halogen changes from I$_i$ to Br$_i$ (and also the instability of +1, when compared to -1, in CsPbCl$_3$) is associated with the increase of the formation energy of the +1 charge state. When the radius of the halogen atom increases, it allows the formation of a trimer *X-X-X* due to the larger space in the faces of the cage that contains the Cs atom. With the decrease of atomic radius (by consequence the area of the cage face), the formation of a trimer of halogens becomes less stable than the formation of two Pb in an environment with seven halogens, as shown in Fig. 3 for CsPbCl$_3$. The consequence in the electronic structure is that the Kohn-Sham (KS) level of the interstitial atom for the +1 charge state is in resonance in conduction band for CsPbI$_3$. When the atomic radius decreases, the KS level for the defect moves inside the band gap for CsPbBr$_3$ (E$_c$ - 0.45 eV) and becomes resonant in the valence band for CsPbCl$_3$. Since the defect level for CsPbCl$_3$ could not be unoccupied in the valence band, this configuration becomes unstable when compared to the negative charge state.

*Compliance vs contradiction of the result for structural defects as dopants with the design principles of ideal doping:* Now that we have identified the lowest formation energy structural defects at any given $E_F$, we can discuss the possibility to dope CsPb$X_3$ (*X* = Cl, Br, I) just by controlling the precursors during the synthesis of the materials, i.e., controlling growth conditions or the chemical potentials. The graphs shown in Figure 4 are valid only for the intermediate halogen condition, corresponding to the point A in the triangles (chemical potential diagrams) shown in Fig. 2. A full analysis, indicating also other chemical potential conditions (halogen rich and poor) are shown in the Supplementary Material (Fig. S2).

We can observe that for all the structural defects studied here, the TTL leads to a shallow donor or acceptor behavior, with exception of I$_i$. If only this DP (i) condition were considered, the doping of both *p*-type and *n*-type should, in principle, be achieved by a simple modification in the precursor ratio during the growth process. The halogen poor condition favors the *n*-type doping while the halogen rich the *p*-type. However, as demonstrated by experimental results, this route of doping has low efficiency, and the increase of the free carrier is very small.[4]

The second DP (ii) is related to the position of the Fermi energy pinning with respect to the band edges.[19,21,42] In our system, CsPb$X_3$ (*X* = Cl, Br, I), $E_{pin}^{(p)}$ (pinning for *p*-type doping) in inorganic HP's is determined at $\Delta H_f(V_X, +) = 0$ for a halogen rich condition, and it lies in the valence band with energy, E$_v$ - 0.11 eV, E$_v$ - 0.56 eV and E$_v$ - 0.47 eV for *X* = I, Br and Cl, respectively. This condition, in addition to the shallow acceptors, indicate that the *p*-type doping could be achieved. On the other hand, $E_{pin}^{(n)}$ (pinning for *n*-type doping) is determined at $\Delta H_f(V_{Pb}, -2) = 0$ for the Pb rich (halogen poor) condition, and lies in the band gap with energy $E_c$ - 0.27 eV and $E_c$ - 0.55 eV for CsPbBr$_3$ and CsPbCl$_3$, respectively, and is located at the $E_c$ edge for CsPbI$_3$ (this can be observed in Fig. S2 for the halogen poor condition). Since the $E_{pin}^{(n)}$ determine the maximum position of the Fermi level, the *n*-type doping will never be achieved for CsPbBr$_3$ and CsPbCl$_3$ because the Fermi level lies deep in the band gap.

Looking for DP (i) and DP (ii), the *n*-type doping could be achieved for CsPbI$_3$, while the *p*-type for all the systems. However we also need to calculate the DP (iii), i.e., the *shift in the doping-induced equilibrium Fermi energy $\Delta E_F^{(n)}$ towards the conduction band for n-type doping and the shift of $\Delta E_F^{(p)}$ towards the valence band for p-type doping.* The equilibrium Fermi level is located close to the crossing point of the donor and acceptor defects with lowest formation energy, as indicated by the dashed line in Fig. 4 and described before. For example, for CsPbI$_3$ the equilibrium Fermi level is close to the crossing of $V_I^{+1}$ and $V_{Pb}^{-2}$, while for CsPbBr$_3$ and CsPbCl$_3$, it occurs at the crossing of $V_X^{+1}$ (*X* = Br, Cl) and $V_{Cs}^{-1}$. In the intermediate condition, Point A in Fig. 2, the equilibrium Fermi level lies in a range $E_v$ + (0.6-0.8) eV for all the halogens, which makes $E_F$ almost in the middle of the band gap for CsPbI$_3$, and between the middle of band gap and $E_v$ for CsPbBr$_3$ and CsPbCl$_3$. Because of DP (i), (ii) and (iii), it is impossible to dope HP by considering only structural defects under the stochiometric growth conditions. If we consider halogen rich and poor conditions, as shown in Fig. S2 in the supplementary material, $E_F$ lies closer to the conduction (halogen poor) or valence band (halogen rich), however it is still located inside the band gap. This explains the reason for a small improvement in the free carriers by controlling the precursors ratio.

One way to shift the equilibrium Fermi level to the CB or VB is by including extrinsic elements. For the *p*-type doping, results shown in the literature indicate that the introduction of acceptor dopants, such as Na, K, Rb, Cu or Ag still keep the equilibrium Fermi level deep inside the band gap.[5,16,30,31] Therefore, the *p*-type doping is impracticable in CsPb$X_3$ (*X* = Cl, Br, I). Here we are going to check only the candidates that lead to a possible *n*-type and the results are shown in the next subsection.

**B. Extrinsic *n*-type impurity dopants**

For the extrinsic doping we analyze only candidates that should be able to lead to a possible *n*-type doping. This includes post transition metals (PTM) such as Al, Ga, In, Sb and Bi; one transition metal, Au; and one rare earth element, La. For all these extrinsic elements we consider only a substitutional site, where the dopant replaces a Pb atom, since the substitution on Cs is less favorable due to the large atomic radius of this element. In addition, to check the possibility of *n*-type doping, we need to consider only the intermediate chemical potential for halogen atoms, point A in Fig. 2. The condition of halogen rich favors the *p*-type doping, as indicated by the position of the equilibrium Fermi level closer to the $E_v$ in Fig. S2 of the supplementary material, and the condition of halogen poor favors the precipitation of the doping precursor as demonstrated in the Theoretical section and disused in the supplementary material.

*Post transition metals, family IIIA*: All the elements studied in this family (Al, Ga, In) show up the +3 formal oxidation state with its valence composed by two electrons in the *s* and one in the *p* orbitals. When this impurity replaces Pb in CsPb$X_3$ (*X* = Cl, Br, I), two electrons tend to be donated to halogen atoms and one unpaired electron in the *s* orbital is observed in the neutral charge state. This configuration with half of the *s* orbitals filled has a very high formation energy, and therefore is less stable than the +1 or -1 charge state, as shown in the formation energy diagrams depicted in Fig. 5 (light blue, green and pink lines). As discussed for $V_X$ and $X_i$, these elements from family IIIA (substitutional at the Pb site) show a negative Coulomb interaction (negative U), i.e., the elements Al, Ga and In tend to form a completely full or empty *s* orbital, and by consequence only the transition from +1 to -1 charge state is observed.

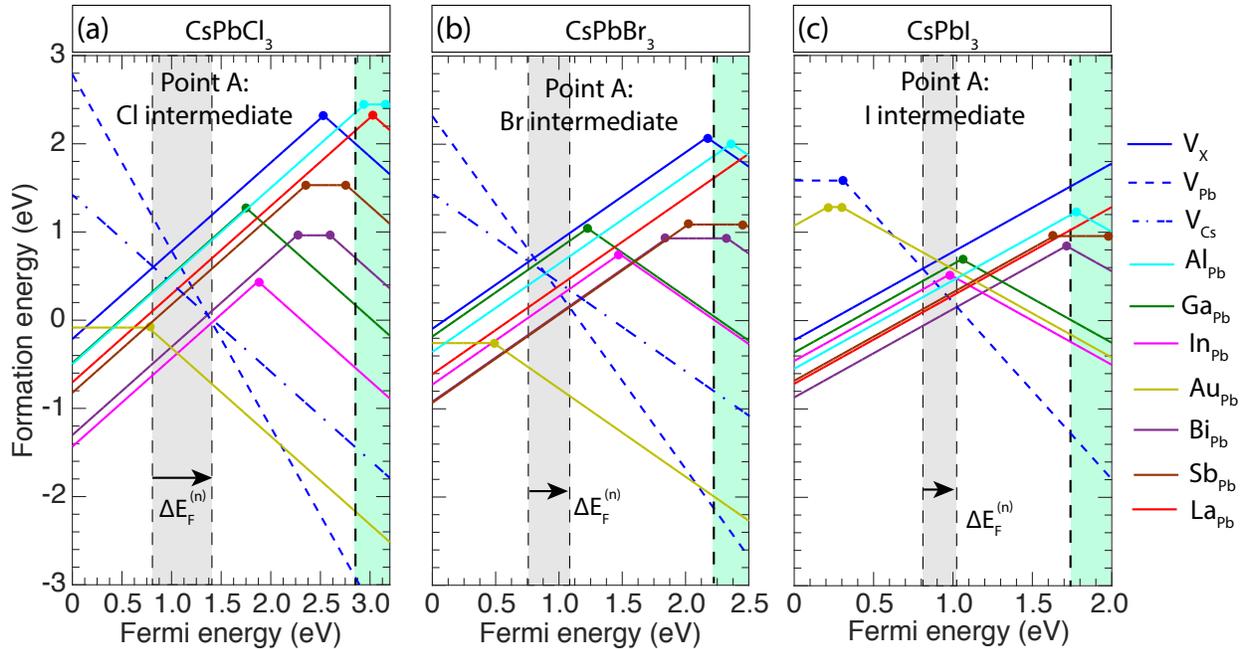

Fig. 5: Formation energy for dopants as function of Fermi level for CsPb$X_3$ with $X$ = (a) Cl, (b) Br and (c) I. The formation energies were calculated only for the point A in the phase diagram of Fig. 2, which represents a stochiometric growth condition, and the halogen chemical potential ensures the introduction of the dopant into the system. The limit for the conduction band is shown by the green area in each system. We also indicate the maximum doping-induced Fermi level shift $\Delta E_F^{(n)}$, considering the crossing of the donor dopant and structural acceptor defects with the lowest formation energies.

For Al doping, the transition (+/-) occurs only inside the conduction band, which makes this element a shallow donor. However, the formation energy of this defect is very high, and comparable to the most stable intrinsic donor defect ($V_X$). Even though the TTL is suitable for a good candidate for *n*-type doping, the large formation energy discards the usability of Al. The incorporation of Ga on Pb site, $Ga_{Pb}$, has a similar formation energy as $Al_{Pb}$ for all the halide perovskites studied here, however the TTL (+/-) occurs deep inside the band gap. This behavior indicates that $Ga_{Pb}$ act as electron compensation defect when the Fermi level lies closer to the $E_c$, and therefore cannot be considered a good candidate for *n*-type. Despite the formation energy of $In_{Pb}$ is smaller than $Ga_{Pb}$, especially in CsPbCl$_3$, the deep TTL inside the band gap leads to the same *n*-type charge compensation of $Ga_{Pb}$ and therefore is considered a poor candidate for *n*-type doping in CsPb$X_3$ ($X$ = Cl, Br, I).

*Post transition metals, family VA*: Sb and Bi have five valence electrons: two electrons in the *s* orbitals and three in the *p* orbitals. This electronic configuration can make Bi and Sb to have two possible formal oxidation states: +5 and +3. When these elements replace Pb in the octahedral site, only the oxidation state of +3 can be observed, thus two valence electrons from the *p* orbital are donated to halogen atoms and only one unpaired electron in *p* orbital (with the two in *s* orbitals) compose the neutral charge condition for Bi and Sb substitutional defects. This configuration has a high formation energy and is more stable than the charged defects for a small window close to the $E_c$ for Sb in all CsPb$X_3$ ($X$ = Cl, Br, I) as shown by the horizontal brown line in Fig. 5. For Bi atom, a negative Coulomb interaction is observed only for CsPbI$_3$, while the formation energy for neutral charge state is the lowest one compared to the charged for $X$ = Cl and Br close to $E_c$. The main reason for the stability of the neutral charge state (one unpaired electron in *p*-orbital) is that the atomic configuration does not change too much when the charge state varies from -1 to 0 and from 0 to +1 with these substitutional atoms. This results in the energy gain for atomic relaxation smaller than the other examples shown here, such as $V_X$ and substitutional atoms from family IIIA.

The formation energy for Bi and Sb in CsPbBr$_3$ and CsPbI$_3$ has a negative value for most available electronic chemical potentials (Fermi energy): from close to E$_v$ until half of the band gap, as shown in Fig. 5. This is a consequence of the good match of the atomic radius of Sb and Bi with Pb, which leads to a small distortion of the local environment and a large similarity to intrinsic CsPbBr$_3$ and CsPbI$_3$. Even when the halogen atomic radius is smaller, such as in CsPbCl$_3$, Bi and Sb show the best results if we do not consider the In substitutional. The charge transition levels (+/-) or (+/0) occurs in a region smaller than E$_c$ - 0.5 eV. From the experimental point of view, Bi can induce a decrease in the resistivity of CsPbBr$_3$.[10]

*Rare earth element*: The substitution of Pb by La leads to a much larger formation energy for neutral and negative charge states if compared to the positive one for all CsPb$X_3$ ($X$ = Cl, Br, I) perovskites. The donation of two valence electrons (one from *d* and one from *s* orbitals) of La to halogen atoms leaves one unpaired electron in the *s* orbital that increase its formation energy, resulting in a decreased stability when compared to the positive configuration. The transition level (+/-) occurs only deep inside the conduction band for all CsPb$X_3$ ($X$ = Cl, Br, I). In the intermediate condition (point A), the formation energy of $\text{La}_{\text{Pb}}^{+1}$ is negative only close to the E$_v$, and is only lower than $\text{Al}_{\text{Pb}}^{+1}$ and $\text{Ga}_{\text{Pb}}^{+1}$. The exception occurs in CsPbI$_3$; the largest volume generated by the I radius increases the stability of $\text{La}_{\text{Pb}}^{+1}$. For CsPbBr$_3$ and CsPbCl$_3$, the large atomic radius of La compared to Pb and the small volume due to the smallest radius of halogen atoms makes this dopant less stable than the others mentioned before, resulting in a bad alternative for *n*-type doping.

*Transition metal*: Au can occur in different formal oxidation states, such as +1, +2 and +3. For this study we expected that Au would assume the +3 formal oxidation state, and act as a *n*-type impurity. However, according to the formation energies shown in Fig. 5, Au acts as an acceptor for most Fermi energies, and therefore it does not assume the +3 oxidation state. For CsPbBr$_3$ and CsPbCl$_3$, the transition level (0/-) is observed with energy E$_v$ + 0.48 and E$_v$ + 0.77 eV, indicating that $\text{Au}_{\text{Pb}}$ is a deep acceptor. In CsPbI$_3$, the $\text{Au}_{\text{Pb}}^{+1}$ has the lowest formation energy close to the E$_v$, the -1 charge state becomes the most stable charge state for $E_F$ 0.30 eV above E$_v$. Different from the other substitutional atoms, $\text{Au}_{\text{Pb}}$ has a negative formation energy for all values of $E_F$ in CsPbCl$_3$ and CsPbBr$_3$, indicating that this impurity is spontaneously incorporated in the material. In CsPbI$_3$, the position of transition level (0/-) for $\text{Au}_{\text{Pb}}$ is more suitable for *p*-type doping. However, the formation energy of this defect is very high, which makes it difficult to incorporate this element. The highest instability of $\text{Au}_{\text{Pb}}$ in CsPbI$_3$, when compared to the other dopants discussed before, is associated to the small atomic radius of Au and the largest cell volume resultant of the I radius.

*Compliance vs contradiction of the result for impurity doping with the design principles of ideal doping:* Similar to the structural defects, the introduction of impurity dopants in CsPb$X_3$ ($X$ = Cl, Br, I) can lead to shallow thermodynamic transition levels and thus fulfilling DP (i), with exception of $\text{Ga}_{\text{Pb}}$ and $\text{In}_{\text{Pb}}$.

As to DP (ii), the Fermi level pinning is a property that depends only on the structural defects, and therefore the introduction of impurity dopants does not change its values. In the previous subsection we demonstrated that $E_{pin}^{(n)}$ lies deep in the band gap for CsPbBr$_3$ and CsPbCl$_3$, demonstrating that the *n*-type doping is hard to be achieved in these perovskites. On the other hand, $E_{pin}^{(n)}$ for CsPbI$_3$ lies in the edge of the E$_c$, and an analysis for the equilibrium Fermi level is needed to verify the possibility for *n*-type doping.

As motioned before, the equilibrium Fermi level is determined by the charge neutrality condition in the system, that takes into account the density of states in the valence and conduction band, the density of structural defects and dopants and their formation energies. The position for $E_F$ is going to be very close to the crossing of donors and acceptors with the lowest formation energies, that could be structural defects or dopants. For all the dopants studied here, the formation energy for $M_{\text{Pb}}^{+1}$ (where $M$ is the dopant) is smaller than halogen vacancy, $V_X^{+1}$ (most stable donor defects among all the structural defects), and therefore the equilibrium Fermi level is determined at the crossing of $M_{\text{Pb}}^{+1}$ and $V_{\text{Pb}}^{-2}$ or $V_{\text{Cs}}^{-1}$. The equilibrium fermi level in the doped system can be compared to the intrinsic one, and the difference in the energy $\Delta E_F^{(n)}$ is the DP (iii) mentioned before and shown in Fig. 1. The maximum $\Delta E_F^{(n)}$ possible, considering all the dopants studied here is shown in the gray area of Fig. 5. For CsPbCl$_3$, In$_{Pb}$ moves $E_F$ by 0.49 eV towards the E$_c$, while in CsPbBr$_3$ and CsPbI$_3$, Bi$_{Pb}$ leads to a shift of 0.28 and 0.22 eV, respectively. For the other possible substitutional atoms, the shift in the Fermi energy is shown in Table S1 in the supplementary material.

The magnitude of $\Delta E_F^{(n)}$ increases if the stability of the donor dopant increases when compared to the most stable structural donor defect, i.e., $V_X^{+1}$ in inorganic HP's. In other words, the smaller the formation energy of the donor dopant, the larger is $\Delta E_F^{(n)}$. Since we demonstrated that most of the donor dopants, e.g. Ga$_{Pb}$, Al$_{Pb}$ and La$_{Pb}$, leads to a high formation energy in the positive charge state, $\Delta E_F^{(n)}$ is small and the equilibrium Fermi level will still lie deep inside the band gap and therefore the DP (iii) is not achieved for *n*-type doping.

Is important to note that the Fermi level after doping is not even close to the pinning of the Fermi level for *n*-type shown in the last subsection. In the condition of maximum pinning of the Fermi level, which is halogen poor, the chemical potential for the halogen favors the precipitation of the dopant precursor, not leading to dopant incorporation. Hence, in this condition we can say that the "effective pinning of the Fermi level" for *n*-type is deeper than the one calculated before, and must lie closer to the middle of the band gap.

Therefore, we demonstrate that, in halide perovskites CsPb$X_3$ (*X* = Cl, Br, I), there are structural defects or dopants that leads to shallow TTL and can successfully meet design principle DP (i). However, we have shown that the pinning of Fermi level, i.e. DP (ii), lies deep in the band gap for CsPbBr$_3$ and CsPbCl$_3$, resulting in a problem to achieve high levels of *n*-type doping in these HP's. Furthermore, the shift in the doping-induced equilibrium Fermi energy $\Delta E_F^{(n)}$ toward the E$_c$ is very small for all the dopants studied here, which results in an equilibrium Fermi level that lie almost in the middle of the band gap and results in an unfavorable condition for *n*-type doping.

If we wish to achieve high levels of doping in a general material we have to consider intrinsic and extrinsic parameters that are associated with our discussed Design principles. In the intrinsic parameters we need to be careful with the composition of the valence and conduction band of the material. For *p*-type doping, it is disable that the valence band shows a not large ionization potential, while for *n*-type doping the conduction band should not have a small electron affinity. When a material does not follow these points, issues are observed such as possibility for charge localization in form of small polarons, and pinning for holes or electrons that lie inside the band gap, DP-(ii). Taking as examples oxides and nitrates, it is known that they are very difficult to be doped as *p*-type because the valence band (composed by O or N p orbitals) lies very deep in energy, which also can induce the formation of small polarons. [22,47–49] For halide perovskites based on Cl, the band gap is wider, leading to a bad position of both valence band and conduction band, and it is the worst scenario between all the HP's discussed.

For the extrinsic parameters (the dopant), it is desirable to have elements that have similar characteristics with the host elements. For example Bi and Sb have a similar atomic radius when compared to Pb, and both atoms like the octahedral environment with the halogens, as observed in their precursors BiX$_3$ and SbX$_3$. Therefore, the introduction of Bi or Sb in the Pb site will not generate a large strain in the system and can lead to a low formation energy of the extrinsic dopants. This is connected to the DP-(i) and (iii), which can shift the doping-induced Fermi level for a large energy showing a shallow TTL. Elements that lead to large distortions because of the small atomic radius, such as Al for *n*-type doping, and Na for *p*-type doping, also show high formation energy and results in a small doping-induced Fermi energy shift with a deep TTL.

Therefore, experiments should analyze the possible contribution of the valence and conduction band in the host system. Anions with high electronic affinity should be avoided for *p*-type doping systems, as well as the elements that contribute to the valence band maximum or conduction band minimum with high localized states (for example d and f orbitals) for both types of doping. This is one possible condition to have the pinning of electrons and holes inside the conduction and valence band, respectively, following the DP-(ii). The dopant introduced in the host material should show similar structural properties, such as ionic radius and similar chemical environment when it is the precursor and the host material. This one possible condition to have a low formation energy that can lead to shallow thermodynamic transition levels, DP-(i), and also a large doping-induced Fermi energy shift, DP-(iii).

## V. CONCLUSIONS

In summary, we demonstrate the bottlenecks for the doping process in halide perovskites (HP's) through the simple discussion of three Design Principles (DP). For effective doping to be achieved, the HP's should follow: (i) *The thermodynamic transition levels $\epsilon(q,q')$ of a dopant must be energetically close (in a range of ~260 meV) to the respective band edges ("shallow level") to assure, at least, a partial ionization under operating temperatures*; (ii) *The n-type (p-type) "Fermi pinning levels" $E_{pin}^{(n)}$ ($E_{pin}^{(p)}$) should be located inside the conduction (valence) band for electrons (holes). Else, doping will be halted by compensation before sufficient carriers are produced.* (iii) *The shift in the doping-induced equilibrium Fermi energy $\Delta E_F^{(n)}$ towards the conduction band for n-type doping (shift of $\Delta E_F^{(p)}$ towards the valence band, for p-type doping) should be sufficiently large for the equilibrium Fermi level to lie close to the conduction (valence) band.* The DP-(i) is easily achieved in HP's by structural defects or extrinsic doping elements for both *p*- and *n*-type doping. DP-(ii) is satisfied only for holes and DP-(iii) fail for both holes and electrons, being the ultimate bottleneck for the *n*-type doping in Iodine HP's. One possible way to overcome these issues and improve the conditions to dope is the passivation of vacancies by, for example, H, or if the ionic conductivity is large enough to compensate the charge killing by the defects. This mechanism can, in principle, shift the pinning of the Fermi level closer to the conduction band or valence band and reduce the charge compensation by intrinsic defects.

## VI. ACKNOWLEDGMENT

The authors thank FAPESP (grants 2019/21656-8 and 17/02317-2) and CNPq for financial support. We also thank the National Laboratory for Scientific Computing (LNCC/MCTI, Brazil) for providing HPC resources on the SDumont supercomputer. Work at CU Boulder on photovoltaic relevant absorption characteristics was supported by the U.S. Department of Energy, Energy Efficiency and Renewable Energy, under the SunShot "Small Innovative Programs in Solar (SIPS)" Project No. DE-EE0007366.